# Should we geoengineer larger ice caps?

Jacob Haqq-Misra

Blue Marble Space Institute of Science
1200 Westlake Ave N Suite 1006, Seattle, Washington 98109, United States
Email: jacob@bmsis.org

**Abstract**
The climate of Earth is susceptible to catastrophes that could threaten the longevity of human civilization. Geoengineering to reduce incoming solar radiation has been suggested as a way to mediate the warming effects of contemporary climate change, but a geoengineering program for thousands of years could also be used to enlarge the size of the polar ice caps and create a permanently cooler climate. Such a large ice cap state would make Earth less susceptible to climate threats and could allow human civilization to survive further into the future than otherwise possible. Intentionally extending Earth's glacial coverage will require uninterrupted commitment to this program for millenia but would ultimately reach a cooler equilibrium state where geoengineering is no longer needed. Whether or not this program is ever attempted, this concept illustrates the need to identify preference among potential climate states to ensure the long-term success of civilization.

**Keywords:** global catastrophic risk; geoengineering; glacial cycles; climate change

## 1. Introduction

Climate catastrophes are events that could compromise the integrity of the biosphere and render the planet unsuitable for human civilization. Anthropogenic climate change has resulted in an increase in greenhouse warming that threatens several potential climate catastrophes. One possibility is the collapse of large ice sheets, such as the Greenland Ice Sheet or the West Antarctic Ice Sheet, which could increase sea level by ten meters or more and cause a shutdown of the thermohaline circulation in the Atlantic ocean (Overpeck et al. 2006; Lenton et al. 2008). Another risk is that a rise in global average temperature of seven degrees Celsius or more will induce heat stress and cause hyperthermia in humans and other mammals (Sherwood and Huber 2010). Even more extreme is the possibility that climate change could initiate a runaway greenhouse state that would lead to the loss of all oceans and leave the planet uninhabitable (Goldblatt and Watson 2012; Ramirez et al. 2014). Any of these events would pose significant challenges to the longevity of human civilization.

Geoengineering provides a possible remedy to these threats by using technology to modify the global climate (Keith 2000). The most direct geoengineering strategy is to reduce the amount of incoming sunlight reaching the surface, which is referred to as "solar radiation management" or SRM (Caldeira et al. 2013)[1]. One of the widely discussed methods for SRM is the injection of reflective particles into the stratosphere (Crutzen 2006; Wigley 2006; Matthews and Caldeira 2007; Rasch et al. 2008; Brovkin et al. 2009; Keith 2010; Goes et al. 2011; Pidgeon et al. 2013) because it provides a "fast, cheap, and easy" (Caldeira 2015) way to cool the planet. A similar effect could be achieved by deploying an orbiting solar shield to reflect away a portion of sunlight from space (Early 1989; Angel 2006), although this option is orders of magnitude more expensive than aerosol injection. Cloud seeding to enhance the reflectivity of marine stratocumulus clouds is another SRM option that can provide significant cooling to particular regions (Jones et al. 2009; Rasch et al. 2009; Korhonen et al. 2010). Other SRM options such as enhancement of oceanic transport or managed changes in the land carbon cycle operate too slowly to be effective at reducing atmospheric carbon dioxide to pre-industrial levels (Lenton and Vaughan 2009).

For present-day climate change, geoengineering can supplement mitigation and adaptation strategies (Keith 2000; Wigley 2006; Caldeira et al. 2013) to avoid crossing dangerous tipping points in the climate that are difficult or impossible to reverse (Lenton 2011). Geoengineering raises significant ethical and political dilemmas (Lovelock 2008; Gardiner 2010; Svoboda et al. 2011; Tuana et al. 2012; Haqq-Misra 2012; Preston 2013; Pidgeon et al. 2013; Svoboda and Irvine 2014) and also makes civilization more vulnerable to other catastrophes (Baum et al. 2013), so geoengineering may serve best as a last resort if all other options fail (Victor et al. 2009; Gardiner 2013). In the distant future, as the sun undergoes its course of stellar evolution, geoengineering may become necessary to prevent the loss of oceans due to a runaway greenhouse and may be the only way of preserving civilization on Earth (Goldblatt and Watson 2012). Whether or not civilization decides to use geoengineering today, it remains a technological option that could be used at some point in the future.

While most geoengineering proposals focus on the short-term use of SRM to alleviate climate threats, another option is to employ a long-term geoengineering strategy to increase the overall climate stability of Earth. In particular, the polar ice caps reflect away a portion of sunlight reaching the surface and help to mediate global temperature. Larger ice caps would serve to reflect away more sunlight and thereby cool the climate, so SRM geoengineering could be used over an extended period of time for the purpose of permanently increasing the size of the polar ice caps. Such a strategy would result in a stable climate with more ice coverage, different from Earth today and with a lower global average temperature.

This possibility of geoengineering to increase the size of the polar ice caps raises a broader issue of deciding exactly which climate states are more or less desirable than others. This paper examines a range of possible stable climate states to consider how humanity could reduce the risk of climate catastrophe through intentional manipulation of the climate system.

---

[1] Other forms of geoengineering, such as carbon dioxide removal (CDR), can also help to address present-day climate concerns by reducing the abundance of greenhouse gases in the atmosphere. This paper will focus on SRM strategies because they offer the possibility of cooling the climate much more than CDR could accomplish (Shaffer 2010; Smith and Torn 2013).

## 2. Hysteresis in climate models

The climate of Earth is powered by the sun, and the amount of sunlight absorbed by the surface is proportional to global average temperature. Of the sunlight that reaches the top of the atmosphere, about 70% is absorbed by the surface of Earth today while the remaining 30% is reflected away by features such as clouds, aerosol, deserts, and glaciers. Planetary albedo refers to the fraction of incoming radiation that is not absorbed by a planetary system, so for Earth today the planetary albedo is about 0.3. In the past, however, Earth has experienced a range of cooler and warmer climate states with more or less ice coverage that corresponds to a larger or smaller planetary albedo. These various climate states have been investigated with computational models to interpret geologic evidence and demonstrate the potential for the Earth system to exist in more than one equilibrium state (e.g., North et al 1981; Haqq-Misra 2014).

The climate system appears to exhibit hysteresis, meaning that a given climate state depends not only on the amount of radiative forcing but also upon its initial conditions. This suggests that more than one stable climate state could exist for a particular level of sunlight and also indicates the existence of irreversible transitions in the amount of glacial coverage. A schematic diagram of hysteresis in Earth's climate is shown in Fig. 1, adapted from a one-dimensional climate model (Haqq-Misra 2014). The horizontal axis corresponds to an increase in radiative forcing toward the right (either from sunlight or greenhouse gases), and the vertical axis indicates the latitudinal extent of glacial ice. Solid lines show stable climate states, while dashed lines show discontinuous transitions between climate states. For many values of radiative forcing there are multiple stable climates, and the availability of a particular climate state depends upon the previous state. In other words, climate evolution is path dependent, so that the future state of global ice coverage is limited by past history.

For example, Earth today has small ice caps on its poles, and an increase in radiative forcing will cause these ice caps to shrink. However, as radiative forcing increases further, these small ice caps eventually reach a critical size where they become unstable and disappear entirely to leave the planet in a completely ice-free state. This transition, known as the "small ice cap instability", is not immediately reversible, and regrowth of the ice caps is only possible by a significant decrease in radiative forcing. In such a scenario, humanity would have to adapt its infrastructure to a warmer planet that absorbs more sunlight with no reflective glacial coverage.

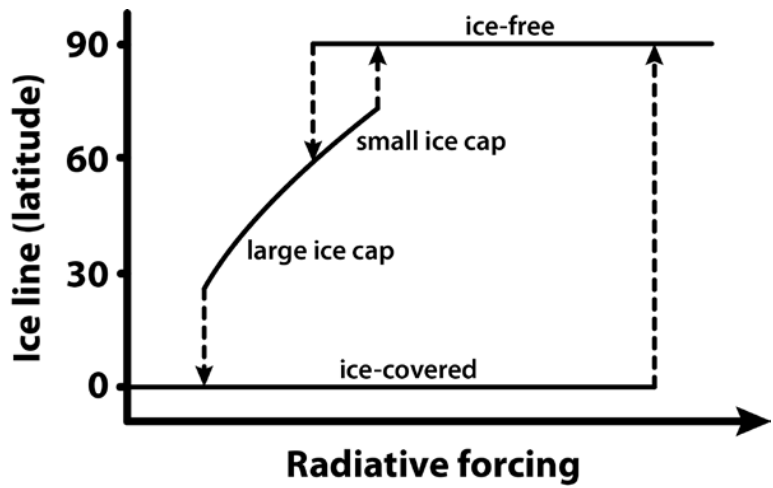

**Figure 1:** Schematic diagram of hysteresis in Earth's climate. Solid lines indicate stable equilibrium climate states with ice-free, small ice cap, large ice cap, and ice-covered conditions. Dashed lines show discontinuous transitions between climate states as radiative forcing increases or decreases.

Model calculations suggest that global average temperature could be anywhere from 4 to 8 K or warmer than today if anthropogenic warming melts the polar ice caps (Lenton and Ciscar 2012; Hansen et al. 2013; Holden et al. 2014; Haqq-Misra 2014). Most climate models today are designed to operate in a regime near present-day Earth, with accuracy up to a few degrees of warming and most of the major ice sheets present in approximately their current configuration. Representation of an ice-free or ice-covered glacial state requires a climate model coupled to a glacial model that can be run for hundreds or thousands of simulated years into the future. Few models are capable of giving reliable predictions for these drastic changes to the Earth system, so the values of temperature change quoted in this paper should be interpreted qualitatively.

If radiative forcing is decreased from the present state, then the cooler climate will yield large ice caps that extend from the poles to as far as about thirty degrees latitude. Calculations suggest that a large ice cap climate could be as much as 10 K colder than today in global average temperature (Caldeira and Kasting 1992; Ishiwatari et al. 2007; Haqq-Misra 2014), depending on the latitudinal extent of the larger ice cap. Such a climate state would require humanity to adapt to a cooler environment with a larger fraction of snow and ice and drastically different weather patterns. This would cause significant displacement of major population centers and agricultural zones, but this would allow Earth to reach a new stable equilibrium that could allow civilization to survive further into the future.

Beyond this latitudinal limit, known as the "large ice cap instability", the planet will plunge into a completely ice-covered state that cannot be reversed easily. Only with a very large increase in radiative forcing can the planet escape from this ice-covered state, which will result in an abrupt transition to an ice-free state. Such a globally glaciated state could reside at 40 K or more below temperatures today (Caldeira and Kasting 1992; DeConto and Pollard 2003; Pierrehumbert 2004; Ishiwatari et al. 2007; Haqq-Misra 2014; Spiegl et al. 2015). Such a climate state would make civilization as we know it impossible and would require a massive shift in infrastructure and a large reduction in global population in order for humanity to survive.

Earth experiences periodic changes in the amount of radiative forcing due to long-term variations in orbital patterns known as Milankovitch cycles. Analysis of ice cores and oxygen isotopes shows evidence of patterns in ice coverage and global temperature with periods of 100,000, 41,000, and 23,000 years that correlate with known variations in the eccentricity, obliquity, and precession of Earth's orbit (Hays et al. 1976; Petit et al. 1999; Jouzel et al. 2007). These periodic ice age cycles indicate transitions between large ice cap, small ice cap, and ice free climate states and show a range in global average temperature of about 10 K. This suggests that the hysteresis found in climate models (outlined by Fig. 1) may be descriptive of Earth's geologic history. In the absence of anthropogenic activity, these ice age cycles due to Milankovitch forcing are likely to continue for the next million years or longer; however, if anthropogenic forcing is large enough, then this could damp or eliminate the ice age cycle altogether (Mitchell 1972; Loutre and Berger 2000; Haqq-Misra 2014; Wolff 2014).

Geologic evidence also exists for the even more extreme case of global glaciation. During the Neoproterozoic Era about 650 million years ago, Earth appears to have undergone a global glaciation event with solid ice or floating glaciers all the way down to the equator (Kirschvink 1992; Hoffman et al. 1998). Global glaciations may have occurred at other times in Earth's history as well in response to abrupt changes that caused drastic cooling on a planetary scale, and the conditions only alleviated when the greenhouse gases from enough volcanic eruptions accumulated in the atmosphere and triggered a massive deglaciation (Caldeira and Kasting 1992; Pierrehumbert 2004). Most remarkable is that life on Earth managed to survive these hostile conditions by adapting to the environmental niches that were available at the time.

The equilibrium climate states described by Fig. 1 highlight more than the planet's geologic history. These four possibilities of a small ice cap, large ice cap, ice-free, and ice-covered state all indicate potential stable climate states that could be goals for a long-term program that seeks to increase climate stability. Geoengineering provides one technological option that could make any of these climate states attainable.

## 3. Geoengineering to increase glacial coverage

Geoengineering through SRM seeks to reduce global average temperature by increasing planetary albedo, which limits the amount of sunlight that reaches the surface. Geoengineering may be ethically problematic as a response to present-day climate change because of the unequal distribution of harms, unforeseen consequences to the environment, and uncertainty regarding governance structures (Lovelock 2008; Gardiner 2010; Svoboda et al. 2011; Tuana et al. 2012; Haqq-Misra 2012; Preston 2013; Pidgeon et al. 2013; Svoboda and Irvine 2014 ). Furthermore, geoengineering requires total commitment in order to succeed, and if a geoengineering program is halted prematurely, then the resulting rapid rise in temperatures could also make civilization more vulnerable to other threats (Matthews and Caldeira 2007; Brovkin et al. 2009; Goes et al. 2011; Baum et al. 2013). While geoengineering could serve as a last resort to prevent climate catastrophe in the near future (Victor et al. 2009; Gardiner 2013), another option with a longer-term vision is to use geoengineering to alter the amount of glacial coverage on Earth as a means of increasing climate stability.

The growth rate of ice sheets depends on environmental factors that allow a net accumulation of moisture. Specifically, the growth of ice sheets is determined by the difference between the rate of mass accumulation by snowfall and the rate of mass loss from melting, sublimation, and wind erosion. Geologic evidence suggests that changes in global temperature from Milankovitch orbital forcing take anywhere from about 6,000 to 10,000 years to cause a corresponding growth or retreat in ice sheet size (Shackleton 2000). However, intentional SRM geoengineering could be used to create strong global cooling that persists for thousands of years and damps out any variations from ice-age cycles. Aerosol SRM geoengineering is likely the cheapest option to enlarge the ice caps (Keith 2010; Caldeira et al. 2013), but a space-based solar shield (Early 1989; Angel 2006) could also provide the needed long-term cooling. Either of these SRM solutions, or a combination of the two, could also be complemented with cloud seeding (Jones et al. 2009; Rasch et al. 2009; Korhonen et al. 2010) and other SRM strategies (Lenton and Vaughan 2009) in order to optimize the growth of the polar ice caps. This would reduce the amount of time required to reach a new climate state, although estimating exact timescales depends on specifics of glacier dynamics and the geoengineering program. Nevertheless, a commitment of a thousand years or more to strong geoengineering could be sufficient to permanently grow the size of the polar ice caps and leave the planet in a significantly cooler state.

The major problem with enacting this proposal is the long-term commitment required to achieve the desired state. Indeed, geoengineering to increase the size of the ice caps cannot alleviate concerns regarding present-day climate change, and the efforts required to sustain such a long-term geoengineering program would be at least as difficult as achieving adequate mitigation among nations. However, this strong geoengineering option to grow the polar ice caps presents a technological solution to increase the long-term climate stability of Earth and reduce the likelihood of potential climate catastrophes. Even so, maintaining a geoengineering program for thousands of years is a challenge that has little precedent in human history, and any failure to keep the program running could set back all process and even initiate a climate catastrophe in the wake of rapid warming. Any long-term geoengineering program to permanently modify the climate will only succeed if structures are in place to ensure the continual pursuit of this goal for an uninterrupted stretch of millenia.

## 4. Selecting a desirable climate state

The prospect of intentionally engineering a more stable climate state raises a deeper issue of what exactly constitutes a more or less desirable climate. Much attention has been given to the regional impacts of climate change scenarios over the coming centuries (IPCC 2014), but little discussion addresses whether or not there are fundamentally better climate states for the purpose of increasing long-term stability.

The present small ice cap state is vulnerable to an increase in greenhouse gases and shows continued signs of warming. If this warming trend continues, then the distant future of Earth's climate could cross the small ice cap instability and result in an equilibrium climate state many degrees warmer than today and with no reflective glacial coverage. Geoengineering in the short term can only provide civilization with extra time to implement mitigation and adaptation measures while simultaneously reducing emissions. Any future civilization in such a climate state will need to depend on fuels that do not further exacerbate greenhouse warming such as renewable sources of energy, nuclear fission, and perhaps nuclear fusion. Failure to address these concerns will risk catastrophic sea level rise (Overpeck et al. 2006; Lenton et al. 2008) or dangerous humidity levels (Sherwood and Huber 2010) that could cripple the foundations of civilization. To this extent, the current small ice cap state may be less than ideal for the long-term future of humanity.

Glacial cycles feature prominently in Earth's geologic record, and the proposal outlined here suggests using a long-term SRM geoengineering program to increase the size of the ice caps permanently. The resulting climate state would be several degrees cooler than today, with the magnitude of cooling dependent upon the goals of the SRM geoengineering program. A range of ice caps larger than today could remain stable (Fig. 1), so this proposal would allow humanity to decide in advance upon a target temperature and ice cap size. Perhaps a modest increase in ice coverage would be desired to counterbalance anthropogenic and environmental variability, or perhaps a larger ice cap would be desired to cool global average temperature by several degrees. This long-term thermostat would provide greater flexibility in the degree to which future civilization can interact with and manipulate the atmosphere. Even if civilization transitions to efficient and renewable sources of energy, rather than emit more greenhouse gases, a large ice cap state would provide a climate that can better withstand changes in the steady brightening of the evolving sun. Mass migrations would be required to shift populations that currently reside in regions that would be covered in ice, likely reducing the total amount of habitable area on Earth. Alterations in land use would be required due to the changes in precipitation and ice cover, which would likely shift agricultural zones and grossly modify regional climate. Yet if the objective is to ensure a stable climate so that civilization can exist for millions of years into the future, then perhaps such changes may be a small price to pay.

An even more extreme possibility is to use geoengineering to intentionally create a globally glaciated state with glacial coverage all the way down to the equator. Such a climate state would demand that humanity is dependent completely upon artificial structures, controlled food production, and efficient energy use in order to withstand the below-freezing climate conditions. The only habitable regions on such a world may be near the equator, requiring a grossly massive relocation reduction of population. Global glaciation would be an unthinkable catastrophe to befall civilization today, but in the distant future such a climate might be the only climate state that can withstand the brightening sun that would otherwise cause a runaway greenhouse and evaporate the oceans (Goldblatt and Watson 2012; Ramirez et al. 2014). This scenario might not be relevant until half a billion years or more into the future, but it is worth remembering that even in these distant situations foresight will be required to avert disaster.

Glacial ice provides a thermostat to Earth's climate, but controlling this thermostat through geoengineering will require dedicated long-term efforts that must remain uninterrupted. This commitment over thousands of years would also require mass relocation of populations, adjustment to drastic shifts in agricultural zones, and redesign of civilizational infrastructure—an expensive endeavor to say the least. The payoff from such an investment will be a more stable climate that can withstand the most extreme threats to climate and will allow Earth's biosphere to thrive for as long as the sun shines.

Might some of these climate states be desirable for the future of civilization? The current small ice cap state is arguably less stable than alternatives, but this is the state in which humanity lives today. This has even prompted the suggestion that coal reserves could be kept on hand for use in the distant future to ward off an oncoming glacial cycle and maintain the current climate state (Stager 2011). Short-sighted decisions that look only to the coming decades or centuries may prefer to maintain the present-day small ice cap state, yet the steady brightening of the sun will challenge this precarious climate and force humanity to consider alternatives. If civilization is to survive for millions, rather than thousands, of years into the future, then some degree of climate manipulation to stabilize a cooler climate may be worth consideration.

## 5. Conclusions

If humanity conquers its current climate challenges, then even greater threats to the biosphere can be overcome through intentionally growing the polar ice caps. Such a feat would require dedicated effort at geoengineering for thousands of years, but this program would help to ensure a stable future for humanity under a cooler climate that is more robust to threats that the climate state today. Such drastic measures to counteract inevitable climate change will become necessary to counteract the brightening of the evolving sun, so even if humanity chooses not to deploy its geoengineering technology today, these same ideas may help someday to extend the ultimate habitable lifetime of the planet.

Regardless of whether or not humanity engages in long-term geoengineering, this proposal illustrates that humanity has already entered the epoch of the Anthropocene, and the impact of civilization upon the global climate is irreversible. The question for the future is not how to undo these actions but how to steer toward a preferable outcome. The ability to manipulate major features of climate is within the grasp of humanity, but humility, rather than hubris, should guide any decisions to control our environment.

**Acknowledgments**
The author thanks Seth Baum and Simon Driscoll for helpful comments and discussions. All opinions are those of the author alone. No funding was used for this research.